\documentclass[8.5pt,twoside]{article}
\usepackage[super,sort&compress,comma]{natbib}

\usepackage{graphicx} 
\usepackage{lastpage}

\begin{document}

\makeatletter \renewcommand{\fnum@figure}{\textbf{Fig.~\thefigure~~}}
\def\subsubsection{\@startsection{subsubsection}{3}{10pt}{-1.25ex plus
    -1ex minus -.1ex}{0ex plus 0ex}{\normalsize\bf}}
\def\paragraph{\@startsection{paragraph}{4}{10pt}{-1.25ex plus -1ex
    minus -.1ex}{0ex plus 0ex}{\normalsize\textit}}
\renewcommand\@biblabel[1]{#1} \renewcommand\@makefntext[1]%
{\noindent\makebox[0pt][r]{\@thefnmark\,}#1} \makeatother


\noindent\LARGE{\textbf{Delayed solidification of soft glasses: New experiments, and a theoretical challenge}}
\vspace{0.6cm}

\noindent\large{\textbf{Yogesh M. Joshi,$^{\ast}$\textit{$^{a}$}
    A. Shahin,\textit{$^{a}$} and Michael
    E. Cates$^{\ast}$\textit{$^{b}$}}}\vspace{0.5cm}



\noindent \normalsize{When subjected to large amplitude oscillatory
  shear stress, aqueous Laponite suspensions show an abrupt
  solidification transition after a long delay time $t_{c}$. We
  measure the dependence of $t_{c}$ on stress amplitude, frequency,
  and on the age-dependent initial loss modulus. At first sight our
  observations appear quantitatively consistent with a simple
  soft-glassy rheology (SGR)-type model, in which barrier crossings by
  mesoscopic elements are purely strain-induced. For a given strain
  amplitude $\gamma_0$ each element can be classified as fluid or
  solid according to whether its local yield strain exceeds
  $\gamma_0$.
  Each cycle, the barrier heights $E$ of yielded elements are
  reassigned according to a fixed prior distribution $\rho(E)$: this
  fixes the per-cycle probability $R(\gamma_0)$ of a fluid elements
  becoming solid. As the fraction of solid elements builds up,
  $\gamma_0$ falls (at constant stress amplitude), so $R(\gamma_0)$
  increases. This positive feedback accounts for the sudden
  solidification after a long delay. The model thus appears to
  directly link macroscopic rheology with mesoscopic barrier height
  statistics: within its precepts, our data point towards a power law
  for $\rho(E)$ rather than the exponential form usually assumed in
  SGR. However, despite this apparent success, closer investigation
  shows that the assumptions of the
  model 
  cannot be reconciled with the extremely large strain amplitudes
  arising in our experiments. The quantitative explanation of delayed
  solidification in Laponite therefore remains an open theoretical
  challenge.}  \vspace{0.5cm}



\footnotetext{\textit{$^{a}$~Department of Chemical Engineering,
    Indian Institute of Technology Kanpur, Kanpur 208016, INDIA. Fax:
    91 512 259 0104; Tel: 91 512 259 7993; E-mail: joshi@iitk.ac.in}}
\footnotetext{\textit{$^{b}$~SUPA, School of Physics \& Astronomy,
    University of Edinburgh, James Clerk Maxwell Building, The King's
    Buildings, Mayfield Road, Edinburgh EH9 3JZ, UK. Fax: 44 131 650
    5902; Tel: 44 131 650 5296; E-mail: mec@ph.ed.ac.uk}}


\section{Introduction}\label{introduction}
Soft materials such as concentrated suspensions, foams, emulsions and
pastes are widely used in products such as foodstuffs, cosmetics,
paints, pharmaceuticals, and ceramic precursors. Many of these systems
show slow dynamics that is attributed to the trapping or jamming of
mesoscopic constituents, creating barriers that the system can cross
only slowly.\cite{cip_co, deb_nat} Such materials can fall out of
thermodynamic equilibrium, evolving by slow ``physical aging" towards
lower energy states, with progressive slowing down of their
relaxational dynamics and rheological response.\cite{str_book,
  bou_jph} Macroscopic sample deformation can in turn promote
barrier-crossing rearrangements, restoring fluidity.\cite{fie_jor,
  fal_pre} The resulting interplay between aging, flow and
rejuvenation in soft glasses leads to complex rheological effects
including overaging, \cite{via_prl, ban_sm} viscosity bifurcations,
\cite{cou_jor, der_epje, chr_pt} and shear banding.\cite{cou_prl,
  var_prl, man_pre, fie_sm}

Here we present new experiments on a model soft glassy material, an
aqueous suspension of Laponite, \cite{sha_lang} under oscillatory
shear of fixed stress amplitude. Building on a preliminary study,
\cite{shu_ces} we find that, despite the high fluidity present
initially, the material abruptly solidifies after a certain critical
time, $t_{c}$. Such ``delayed solidification" could have serious
consequences if it arose, for instance, during the sustained vibratory
stresses that arises during transportation (e.g., in road tankers) of
soft materials, or in some cases during their manufacture. The sudden
solidification of a supposedly fluid formulation risks catastrophic
failure of expensive equipment -- a situation comparable to the issue
of silo rupture, caused when granular materials cease suddenly to
flow.\cite{silos} The problem is all the more serious because of its
apparent unpredictability: nothing obvious about the initial sample
indicates that its fluidity will later be lost in this
way. \cite{shu_ces} The phenomenon of delayed solidification is
thereby reminiscent of (and yet almost opposite to) that of ``delayed
sedimentation" in which an apparently stable colloidal gel suddenly
collapses after sustained exposure to gravitational
stress.\cite{del_sed} There are some reports in the literature where
different soft glassy materials have been observed to undergo delayed
solidification in creep flow.\cite{cou_jor, chr_pt, erw_jor} The major
difference between these reports and our work is the nature of applied
deformation field. We employ a stress-controlled oscillatory flow
field, in which the strain induced in the material remains bounded
(and therefore decreases with increasing viscosity and/or
elasticity). On the other hand, in creep flow the viscous strain
induced in the material never decreases with time.

In this Discussion Paper we attempt to shed light on the physics of
delayed solidification, by performing new quantitative experiments to
explore the dependence of the delay time $t_{c}$ on the stress
amplitude, frequency, and the (age-related) loss modulus of the sample
prior to the oscillatory stress being applied. Alongside this we
develop a simple yet semi-quantitative theory that apparently relates
these dependences directly to the distribution of energy well depths
in soft glasses. Initially we will present this theory at face value,
in tandem with our new experimental results (Sections \ref{main1} and
\ref{main2}). However, in Section \ref{problem} we will show that the
model's assumptions cannot be reconciled with the extremely large
strain amplitudes that arise in our experiments throughout the
prolonged delay period prior to final solidification. In Section
\ref{food} we discuss delayed solidification in the context of food
materials. Our reluctant conclusion (Section \ref{conclusion}) is
that, at least in its present form, our theory is not quantitatively
credible in the context of the Laponite system studied here, although
it may well be applicable to delayed solidification in other soft
glasses. Thus our data poses an open challenge: to create a consistent
quantitative theory of delayed solidification in Laponite.

\section{Experimental methods and results} \label{main1}

In this work we have used aqueous suspensions of 3.2 and 2.8 wt.\%
Laponite RD$^{\textregistered}$ (Southern Clay Products, Inc.),
prepared as described elsewhere. \cite{sha_lang} Each Laponite
suspension was stored in a sealed polypropylene bottle at 30$^\circ$C
for a predetermined ``idle time" $t_{i}$ (7-28 days), then loaded in a
Couette cell (bob diameter 28 mm, gap 1 mm) of an AR1000 stress
controlled rheometer, and shear-melted for 15 minutes by applying
oscillatory shear stress of amplitude 80 Pa at frequency 0.1
Hz. Immediately after this shear melting, an oscillatory stress of
amplitude (in the range 5 - 40 Pa) and frequency $f$ was applied, and
the strain evolution was recorded. Experiments were performed at
25$^\circ$C; the free surface of the sample was covered with a thin
layer of inviscid silicone oil to avoid water evaporation.

\begin{figure}[tb]
  \centering
  \includegraphics[width=4in]{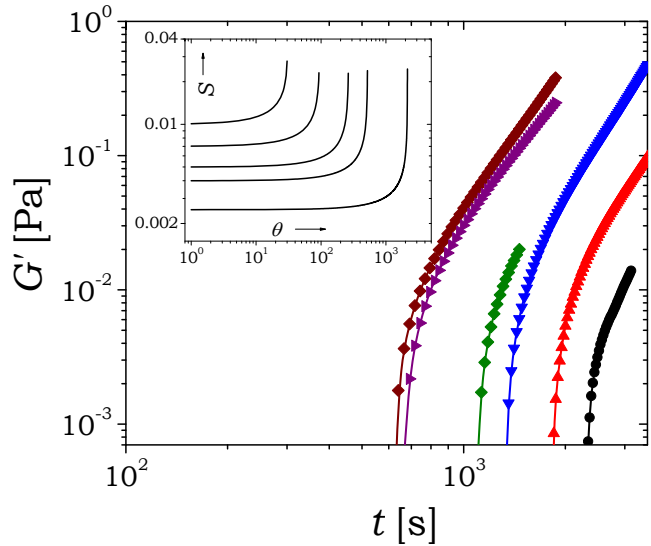}
  \caption{Evolution of $G'$ for various idle times. Evolution of $G'$
    subsequent to shear melting as a function of time for various idle
    times (from right to left: $t_{i}$=7, 10, 13, 15, 24, 28 days)
    under application of oscillatory shear stress of 30 Pa at 0.1 Hz
    frequency for 3.2 weight \% suspension. Inset shows prediction of
    Equations (1-3) for $\lambda/E_{min}$=0.01 and $y$=3, with various
    initial solid fractions $S(0)$.}
  \label{fgr:figure1}
\end{figure}

Our Laponite suspensions generally have a paste-like
consistency. Aging in these systems increases both the elasticity and
the relaxation time. Figure~\ref{fgr:figure1} shows the elastic
modulus $G'$ as a function of time elapsed (after the shear-melting
pre-treatment ends) for experiments at various idle times. Although
the strain induced in the material is large, the characterization of
the aging data in terms of the linear storage modulus $G'$ in
figure~\ref{fgr:figure1} is justified, as the third strain harmonic is
less than 20 \% of the first.\cite{shu_ces} The modulus $G'$,
initially too small to detect, shows a sudden and dramatic increase
(delayed solidification) beyond a critical time $t_{c}$. This arises
despite the very large fluidizing strains, in the range 10-200,
experienced by the material in the first few (post-shear-melting)
cycles.

\begin{figure}[h]
  \centering
  \includegraphics[width=4in]{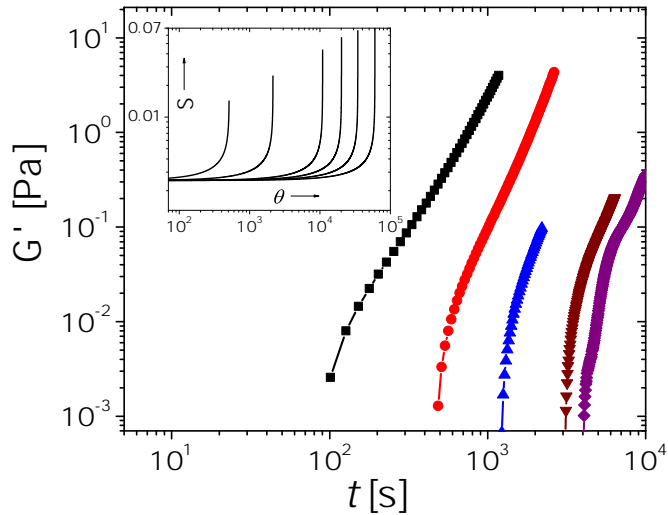}
  \caption{Evolution of $G'$ at various stresses. Evolution of $G'$
    with respect to time under oscillatory flow field having various
    magnitudes of stresses (from left to right: 20, 25, 30, 35, 40 Pa)
    at 0.1 Hz frequency for idle time $t_{i}$= 13 days for 3.2 weight
    \% suspension. Inset shows prediction of Equations (1-3) for $y$=3
    (from left to right: $\lambda/E_{min}$=0.005, 0.01, 0.023, 0.031,
    0.04, 0.053).}
  \label{fgr:figure2}
\end{figure}

\begin{figure}[h]
  \centering
  \includegraphics[width=4in]{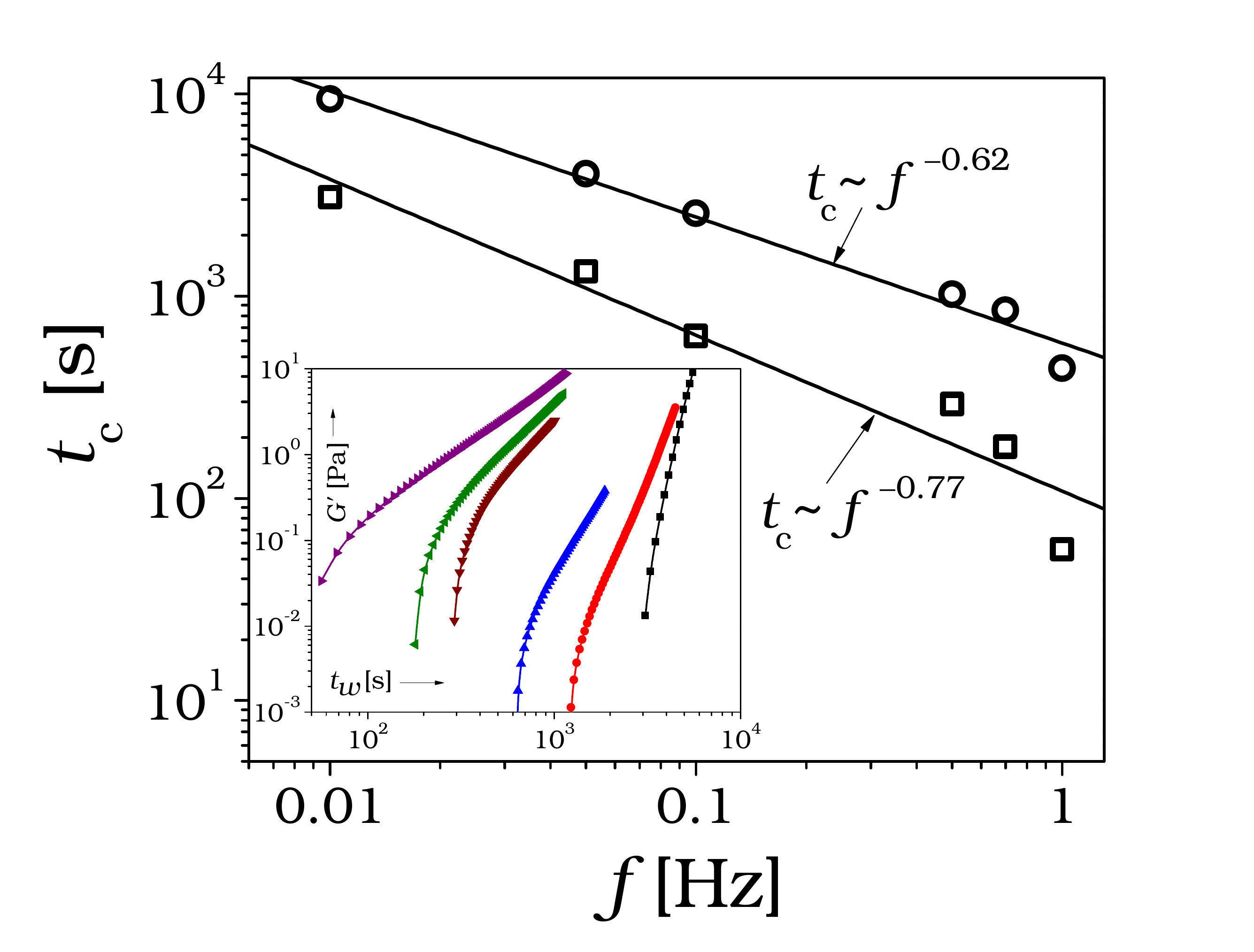}
  \caption{Evolution of $G'$ at various frequencies and dependence of
    critical time on frequency. The critical time for delayed
    solidification plotted against frequency of oscillations (open
    squares: 3.2 weight \% suspension, $t_{i}$=28 day, $\sigma_{0}$=
    30 Pa; open circles: 2.8 weight \% suspension, $t_{i}$=21 day,
    $\sigma_{0}$= 20 Pa). Inset shows corresponding evolution of $G'$
    as a function of time for various frequencies (from right to left:
    0.01, 0.05, 0.1, 0.5, 0.7, 1 Hz) for 3.2 weight \% suspension.}
  \label{fgr:figure3}
\end{figure}

Interestingly, $t_{c}$ decreases markedly as the idle time is
raised. Thus, in contrast to many soft glasses, the aging of Laponite
is partly irreversible: subsequent shear melting does not rejuvenate
all samples to the same state. \cite{sha_lang} Although the
microscopic details of this are debatable,\cite{mon_jcis, ruz_sm} an
aging Laponite suspension at rest clearly crosses some barriers too
high to be reversed by our shear-melting protocol, allowing faster
delayed solidification in older samples
(figure~\ref{fgr:figure1}). However, the evolution of $G'$ after
$t_{c}$ is similar in all cases, suggesting that idle time and aging
time are partly interchangeable. \cite{sha_lang}
Figure~\ref{fgr:figure2} shows how the evolution of $G'$ depends on
$\sigma$ at fixed idle time $t_{i}$. In this set of experiments each
initial state is completely equivalent before the final oscillatory
stress is applied. We see that $t_{c}$ gets larger as the stress is
increased. In addition, Figure~\ref{fgr:figure3} shows that $t_{c}$ is
reduced as $f$ is raised.

\section{A simple model and its predictions}\label{main2}

The microstructure of Laponite suspensions is variously argued to be a
repulsive glass or an attractive gel; \cite{mon_jcis, ruz_sm} under
our conditions (without added salt), repulsions and attractions
probably both are important. \cite{sha_lang} Nonetheless, whether
caging or bonding dominates locally, mesoscopic elements can be
considered trapped in energy wells of various depths. These elements
are forced out of their traps by macroscopic deformation, whereupon
they form new cages or bonds that in turn present new barriers to
rearrangement.

Our model for this process is essentially a soft-glassy-rheology (SGR)
model, \cite {fie_jor, sol_prl, sol_pre,via_prl} considered for
simplicity in the noise-free limit whereby elastic elements cross
barriers only when their local mechanical yield threshold is exceeded.
One important simplification within the SGR approach, which our model
inherits, is the assumption that all elements strain affinely with the
imposed flow between one jump and the next. Allied with the further
simplifying assumptions of harmonicity within each trap and a uniform
elastic constant for all traps, \cite{sol_prl} this represents a
picture in which the intra-jump elastic deformation is that of a
parallel mechanical circuit. The opposite assumption would be to
suppose equal stress in all elements, \emph{i.e.}, a series mechanical
circuit. The real distribution of local elastic deformations and
stresses must lie somewhere between these extremes; we return to this
point in Section \ref{problem}.

Within the SGR framework, the arrested state (an amorphous solid) of a
soft glass is described, as indicated above, in terms of mesoscopic
elements trapped in energy wells created by neighbors. \cite{fie_jor}
A crucial postulate of SGR is that these well depths (or barrier
heights) are distributed broadly. As shown by Bouchaud, \cite{bou_jph}
if the a priori (prior) distribution of well-depths varies as
$\rho(E)\sim\exp[-E/\langle E \rangle]$, ergodicity is lost at a glass
transition temperature $T_{g}$, obeying $k_{B}T_{g}=\langle E
\rangle$. All other forms of $\rho({E})$ lie either in the glass or
the fluid according to whether their decay is slower or faster than
exponential. The SGR model \cite{sol_prl, sol_pre} further allows for
deformations, and replaces the thermal energy $k_{B}T$ by a
nonequilibrium noise amplitude. Assuming exponential $\rho(E)$, the
SGR model offers a unified phenomenological model of soft-glass
rheology, whose predictions include power-law fluid and
Herschel-Bulkley behaviours.  \cite{fie_jor}

Although theoretical arguments suggest it asymptotically,
\cite{bou_jph, sol_pre} there is so far no direct experimental test of
whether the well-depth distribution in soft glasses is indeed
exponential, as SGR assumes. (Other forms might still support an
arrest transition, but only if more complicated cooperative dynamics
are considered. \cite{sol_prl, sol_pre}) It would therefore be useful
to gain clearer experimental insight into the true form of $\rho(E)$.

The prior distribution of well depths, $\rho(E)$, is not their
occupancy probability $P(E)$ since deeper wells have lower escape
rates and are more likely to be occupied. However $\rho(E)$ is the
distribution from which $E$ is drawn once a rearrangement is made and
a new barrier height chosen. If oscillatory shear of amplitude
$\sigma$ is imposed, creating a strain amplitude $\gamma_{0}$, then
assuming affine intra-jump deformation (as SGR does) each individual
element will gain an energy $k\gamma_{0}^{2}/2 \equiv E_{0}$, where
$k$ is a spring constant, at the extremes of each cycle. (As in SGR,
$k$ is here taken independent of well-depth $E$; \cite{fie_jor} we
partially relax this assumption below.) We assume that any element of
well-depth $E<E_{0}$ is rejuvenated during the given strain cycle,
while all those occupying deeper wells are not.

We now distinguish a liquid fraction $F$ and solid fraction $S=1-F$,
representing in turn the fractions of elements occupying wells
shallower or deeper than $E_{0}$. Taking for convenience a time
coordinate $\theta=ft$, we propose $F$ to obey:
\begin{equation}
  \frac{dF}{d\theta}=-FR
\end{equation}
where $R$ is the fraction of jumps into the solid state:
\begin{equation}
  R=\frac{\int\limits_{E_{0}(\gamma_{0})} ^\infty \rho (E)\,\mathrm{d}E}{\int\limits_{0} ^\infty \rho (E)\,\mathrm{d}E}
\end{equation}

Equation 1 models the random events by which `fluid' elements ---
those that cross their barriers by strain-induced dynamics during one
cycle --- become `solid' elements (those that don't) in the next. In a
material having solid fraction $S$ and thus modulus $GS$, the strain
induced by the stress of amplitude $\sigma$ is then taken to be:
\begin{equation}
  \gamma_{0}=\frac{\sigma}{SG}
\end{equation}
The threshold energy at given stress $\sigma$ and solid fraction $S$
then obeys $k\gamma_{0}^2/2 = E_{0}$, or equivalently $E_{0}=\lambda /
S^2$ where $2\lambda = k\sigma^2 / G^2$.

Equations (1-3) can now be solved to give $S(\theta)$ in terms of
$\rho(E_{0})$ and the initial solid content which we denote as
$S(0)=\epsilon$. We consider two cases: first the exponential
distribution $\rho(E)=e^{-\alpha E} / \alpha$ for which
$R(E_{0})=\epsilon^{-\alpha E_{0}}$, and secondly a power law,
$\rho(E)=AE^{-y}$ for $E>E_{min}$, with a cutoff $E_{min}$. The cutoff
is defined so that $\rho(E<E_{min})=0$. By normalization we then have
$A=E_{min}^{1-y}/(1-y)$; the resulting $R$ is given by
$R(E_{0})=(E_{0}/ E_{min})^{1-y}$.

Figure~\ref{fgr:figure1} (inset) shows for the power-law case the time
evolution of the solid fraction $S(\theta)$, for various initial solid
contents $\epsilon$, but the same values of $\lambda/E_{min}$
(equivalently, the same stress $\sigma$). Our simple model captures
both the sudden solidification at $t_{c}$, and the decrease in $t_{c}$
with increasing initial solid content. Figure~\ref{fgr:figure2}
(inset) shows $S(\theta)$ for various values of $\lambda/E_{min}$ (or
equivalently various $\sigma$) but the same initial solid content
$\epsilon$. As observed experimentally, the delayed solidification
time $t_{c}$ is predicted to increase with stress. Since $\theta=ft$,
Equation (1) directly implies $t_{c} \propto
f^{-1}$. Figure~\ref{fgr:figure3} (inset) shows a decreasing trend,
although the best fit power laws (albeit with limited data) are $t_{c}
\propto f^{-0.77}$ for the 3.2\% sample and $t_{c} \propto f^{-0.62}$
at 2.8\%. Given the simplicity of the model, these are close enough to
the $f^{-1}$ prediction to be broadly encouraging.

\begin{figure}[tb]
  \centering
  \includegraphics[width=4in]{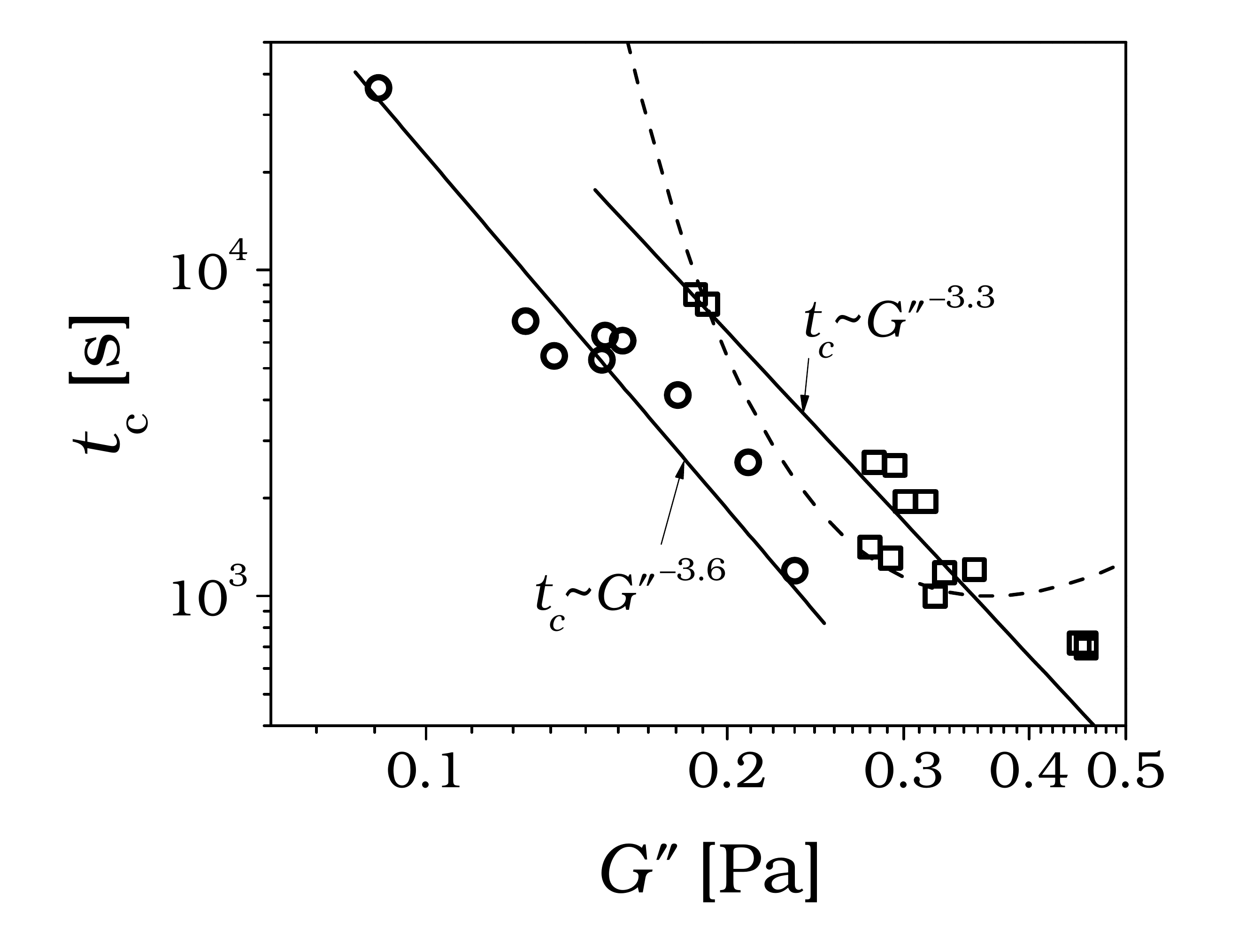}
  \caption{Dependence of critical time on initial $G''$. The critical
    time for delayed solidification plotted against $G''$ measured
    directly after the shear-melting protocol is complete (open
    squares: 3.2 weight \% suspension, $\sigma$= 30 Pa, $f$=0.1 Hz;
    open circles: 2.8 weight \% suspension, $\sigma$= 20 Pa, $f$=0.1
    Hz). Solid lines represent fit of equation (7) while dashed line
    represents fit to equation (6) with $G'' \propto \epsilon$ assumed
    in both cases. The stated exponents for the modulus dependence
    translate into the $y$ values quoted in the text.}
  \label{fgr:figure4}
\end{figure}

Moreover, integrating Equation (1) gives the time at which a given
solid fraction is attained:
\begin{equation}
  \theta(S)=\int\limits_{\epsilon} ^{S} \frac{1}{(1-S')R(S')}\,\mathrm{d}S'\nonumber
\end{equation}
The critical value $\theta_{c}$ identifies the time where $S$ ``ceases
to be small". The precise definition of this quantity is clearly
somewhat arbitrary, but once it is no longer small, $S(\theta)$
increases so steeply that the details of the definition barely
matter. Thus, it suffices to ignore the saturation that occurs as $S
\rightarrow 1$ in the denominator, and then set $S=1$ as the upper
limit of the integral above; these two simplifications give the
following expression for the solidification time $\theta_{c} =
ft_{c}$:
\begin{equation}
  \theta_{c} \approx \int\limits_{\epsilon} ^{1} \frac{1}{R(S')}\,\mathrm{d}S'
\end{equation}
For our exponential and power law distributions the results are
respectively:
\begin{equation}
  \theta_{c}=\frac{2\epsilon^{3}G^{2}}{\alpha k \sigma^{2}} \exp[\frac{\alpha k \sigma^{2}}{2\epsilon^{2}G^{2}}]
\end{equation}
\begin{equation}
  \theta_{c}  \propto \epsilon^{3-2y} \sigma^{2(y-1)}
\end{equation}
Figure~\ref{fgr:figure4} shows $t_{c}$, now experimentally defined by
$G'(t_{c})$ = 5 mPa, as a function of $G''$, the initial loss modulus
measured directly after shear melting. Strictly speaking, our model
relates the solid fraction only to $G'$, which is too small to measure
in this early time regime. Therefore we assume that the loss modulus
is a similar indicator of solid content: $G'' \propto \epsilon =
S(0)$. If so, Equation 6 demonstrates a poor fit to the experimental
data 
but the power law result (Equation 7) fits well, with
$y$=3.15$\pm$0.13 at 3.2\% and $y$= 3.3$\pm$0.18 at 2.8\%.

\begin{figure}[tb]
  \centering
  \includegraphics[width=4in]{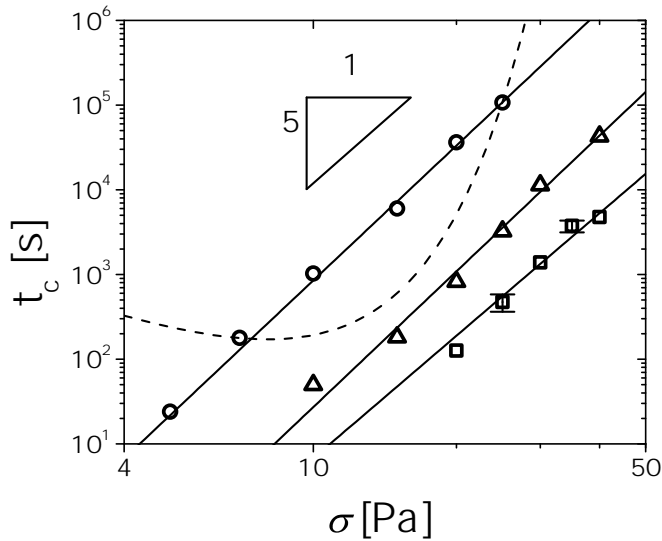}
  \caption{Dependence of critical time on stress. The critical time
    for delayed solidification plotted as a function of applied stress
    amplitude $\sigma$ (circles: 2.8 weight \% suspension, $t_{i}$=9
    day, $f$=0.1 Hz; triangles: 2.8 weight \% suspension, $t_{i}$=15
    day, $f$=0.1 Hz (data from ref: \cite{shu_ces}), squares: 3.2
    weight \% suspension, $t_{i}$==13 day, $f$=0.1 Hz.) Solid lines
    represents fit of equation (7) to the experimental data; the
    dashed line is a fit of equation (6) for the uppermost dataset
    only.}
  \label{fgr:figure5}
\end{figure}

Figure~\ref{fgr:figure5} shows how $t_{c}$ depends on $\sigma$ for the
3.2\% data of figure~\ref{fgr:figure2}, and for two datasets at
2.8\%. Here stress was varied at fixed idle time, so we expect
$\epsilon$ to be fixed, giving $t_{c} \sim \sigma^{2(y-1)}$ by
Equation (7). (Note however that the measured initial $G''$ decreases
with $\sigma$; see the Appendix for a discussion.)
Figure~\ref{fgr:figure5} confirms this power law, and again contrasts
with the prediction for exponential $\rho(E)$. A fit gives $y$
=3.4$\pm$0.3 for the 3.2\% sample and $y$ =3.6$\pm$0.1 for both 2.8\%
samples.

To summarize, fitting separately the dependences of $t_{c}$ on initial
solid content and on stress leads to very similar $y$ values for three
different samples, suggesting a fairly robust power law. This is
interesting from a glass physics viewpoint: any power law distribution
should, by the arguments of Bouchaud, \cite{bou_jph} lie deep in the
aging glass regime. In the Appendix, we generalize our simple model to
allow the elastic constant $k$ of an element to vary as $k \propto
E^{p}$. This gives the same results as above, but with $y$ replaced by
$y_{eff}=(y+p)/(1+p)$; the same conclusion applies.  Thus we have
found that, when interpreted within the precepts of the SGR-inspired
model presented above, our quantitative delayed solidification
measurements for Laponite indicate a power law distribution of the
energy well depths. This apparently direct connection between
macroscopic rheological observations and the mesoscopic energetics is
tantalizing: it holds out the prospect for ``spectroscopic'' analysis
of the energy landscape through careful study of the nonlinear
rheology. In turn this might provide new insight into aging and
rejuvenation in this important class of materials.

\section{What's wrong with this picture?}\label{problem}
Before answering this question, it is interesting to speculate where
power laws in the local yield energy and/or elastic constant might
come from. At the low volume fractions present in our Laponite
samples, bonding might lead to some sort of percolation
transition. Near such a transition the elastic elements comprise
clusters of all sizes, whose moduli are controlled by power laws
(coinciding in the scalar limit with those for
resistivity. \cite{sta_pr}) The corresponding yield energy
distribution is model-dependent, but seemingly can itself exhibit
power laws over one or two decades, or more in some limits,
\cite{str_pre, hun_tpm} although not enough is yet known to suggest
specific values for the relevant exponents.

However the emerging picture of a buildup of the solid fraction $S$
(from an initially minimal level) by a percolation-like process gives
pause for thought. For if solidification is initiated by the formation
of relatively rigid clusters within a sea of fluidized material, SGR's
assumption of a parallel mechanical circuit becomes highly
suspect. One could not expect such aggregates to deform affinely under
any type of flow: solid objects floating in a fluid develop only small
deformations before achieving stresses that match those of their
continuously deformable surroundings. At first sight, this objection
might not appear fatal to our model since, in its many other
predictions, the SGR approach is empirically successful although in
practice local deformations are never affine. Nonetheless, the
percolation viewpoint argues for a model that lies much nearer the
series-circuit (equal stress on all elements) end of the spectrum than
the parallel-circuit assumption embodied in SGR.

A serious blow is struck by noting that, in our Laponite studies, the
enormous initial strains to which the initially fluidized sample is
subjected (of order 10-200, \emph{i.e.,} 1000\% to 20,000\%) are
maintained almost throughout the incubation period prior to the final
solidification event. It is scarcely credible that \emph{any}
mesoscopic element of the type envisaged by SGR could have high enough
internal or external energy barriers to sustain an affine deformation
of this magnitude. By completely destroying the structures that SGR
requires to survive from one cycle to the next, such strain amplitudes
preclude the slow buildup of a solid component which is an essential
precursor to the final dramatic solidification. (Recall that the
latter occurs when the feedback between the slowly building elasticity
and decreasing strain amplitude finally takes over.) The observation
of delayed solidification at these initial strain amplitudes in
Laponite therefore means we must look for a mechanism involving
pockets of solidity that do not deform affinely, and, for that reason
alone, can grow from one cycle to the next. Whatever its merits in
other context, by assuming affine deformation, SGR precludes a
consistent description of any such mechanism.

We can however, speculate a mechanism by assuming that the solid
pockets, and the fluidized suspension surrounding these, share the
same stress (series mechanical circuit). Consequently the strain
induced in the solid region would be very small. Such a scenario may
give rise to an apparent boundary layer around the solid pocket,
wherein strain magnitude changes from practically zero at the surface
to a very large value away from it. In this small strain region very
near the solid surface, the liquid suspension may undergo aging
following a very similar dynamics mentioned in the previous
section. Owing to this aging of the liquid suspension near the solid
surface, the solid region is expected to grow, which in turn will
enhance the fraction of liquid suspension undergoing smaller
strain. Moreover the enhanced solid fraction will also reduce the bulk
strain magnitude. Through a forward feedback mechanism, the growing
solids will fill the space causing jamming of the system as a
whole. This picture preserves some of the physical features of our
SGR-based approach, but would require a different mathematical
description from the one we present above.

\section{Delayed solidification in the context of food
  materials}\label{food}

Interestingly, there are many food materials that have paste like
consistency and are expected to demonstrate soft glassy rheological
behavior. These include: fruit jams, mustard, jellies, mayonnaise,
cheese, ice-creams, tomato and chocolate puree, toothpaste (not
exactly a food material, but edible), etc..  Among these both
mustard\cite{cou_jor06} and mayonnaise\cite{cru_pre} have been
reported to demonstrate physical aging (time dependent enhancement in
viscosity and elastic modulus), which is a signature of soft glassy
behavior. Rheological behavior is one of the most important
characteristic features of food materials. The effect of time and
deformation, which respectively tend to enhance and reduce viscosity
and elasticity, is a very important consideration when designing food
processing equipment and determining the shelf life of a food
product. The present work suggests that under application of a
sustained oscillatory deformation field, food materials may in some
cases transform from an apparently fluid like state to solid
state. Such materials are subjected various kinds of deformation
fields during preparation, transportation and handling. The present
work suggests that such deformation fields can delay but may not stop
the process of aging which leads to solidification.

In addition to physical aging, which is a reversible process, food
materials are also prone to undergo partly irreversible changes in
their rheological behavior. Particularly enzymes, heating or
acidification can induce gelation in certain food products causing an
increase in viscosity and elastic modulus. Thus flowing liquids get
converted to soft solids as a function of time.\cite{eri_fh} The
solidification that we discuss in this manuscript is essentially
reversible. However, Laponite suspensions are also known to show
partial irreversible aging behavior and therefore interestingly mirror
what happens in certain types of food materials.

Besides time-dependent irreversible phenomena, irreversible
aggregation induced by an applied deformation field is also possible
and is particularly observed for proteins. It is known that mis-folded
proteins tend to aggregate because of inter molecular hydrophobic
associations.\cite{oat_rsi} In addition, one of the proposed
mechanisms for spider silk formation also suggests deformation induced
self-assembly of proteins which is irreversible in
nature.\cite{lele_ces, jin_nat, hol_sm} In colloidal suspensions,
shear-induced aggregation has also been reported.\cite{zac_prl}
Interestingly, shear-induced enhancement in elasticity is also
observed in some soft glassy materials\cite{via_prl} and in Laponite
suspensions in particular.\cite{ban_sm} However, there seems to be no
direct connection between shear-induced solidification (for instance
in steady shear) and the delayed solidification addressed in our work
under sustained oscillatory shearing.

\section{Conclusions}\label{conclusion}
We have studied in detail the sudden and dramatic enhancement in
elastic modulus at late times seen in aqueous Laponite suspensions
undergoing stress-controlled oscillatory shear. The critical time
$t_{c}$ for this delayed solidification is reduced for older samples,
despite our use of a vigorous pre-shear protocol, whereas application
of higher stress amplitude, or lower frequency, increases $t_{c}$. We
have proposed a simple SGR-type model wherein a liquid fraction of
fluidized elements are rejuvenated every cycle. At each such event,
there is a probability $R$ of jumping into a solid fraction of deep
wells, that do not rejuvenate; $R$ is determined by the current strain
amplitude, and the prior distribution of well depths. The
ever-increasing solid content slowly decreases the induced strain and
increases $R$; this positive feedback leads eventually to sudden
jamming of the whole sample. Taken at face value, the model offers a
semi-quantitative explanation of our Laponite experiments, with strong
evidence that a power law distribution of energy well depths must be
chosen in preference to the exponential form normally adopted in an
SGR context.  This offers a tantalizing glimpse of how macroscopic
rheology might be directly relatable to the barrier distribution for
rearrangements --- a quantity that has previously eluded direct
experimental characterization in soft glasses. But unfortunately this
is no more than a glimpse, because on closer inspection the large
strain amplitudes arising in the Laponite system cannot credibly be
reconciled with one of the model's central approximations: that local
deformations between rearrangement events follow affinely the
macroscopic flow.

Where does this critique leave the model? Unless its foundations can
be reinterpreted or repaired (and we have not managed to do this so
far), then despite its semi-quantitative success at fitting the data,
our SGR-based approach clearly does not offer a secure starting point
to understand delayed solidification in Laponite
suspensions. Accordingly, the evidence that it seemingly offers for a
power-law rather than exponential barrier height distribution must now
be set aside.

On the other hand, the qualitative physical predictions of our
SGR-inspired model remain intact for systems that are credibly
approximated by its assumption of affine deformation between
rearrangements. We can see nothing to prevent the existence of soft
glasses for which a delayed solidification scenario arises at much
more modest strains, of order unity: our model, even if it must be
rejected for Laponite, offers a ready-made description for such
systems. Meanwhile it remains an open theoretical challenge to develop
a more suitable quantitative model for delayed solidification in
Laponite itself. Only when such a model exists can one know how much,
if any, of the qualitative physics embodied in our model is also
relevant to the Laponite system.

\textbf{Acknowledgements:} We thank the Royal Society of Edinburgh ---
Indian National Science Academy International Exchange Programme for
sponsoring YMJ's visit to Edinburgh (2010). YMJ also acknowledges the
IRHPA scheme of the Department of Science and Technology, Government
of India. MEC is supported by the Royal Society, by EPSRC/EP/EO30173,
and by EPSRC/EP/J007404; he thanks KITP Santa Barbara for hospitality,
where this research was supported in part by the National Science
Foundation (USA) under Grant no. NSF PHY05-51164.

\section*{Appendix}

\subsection*{A. Stress dependence of viscous modulus at fixed idle
  time}

\begin{figure}[h]
  \centering
  \includegraphics[width=4in]{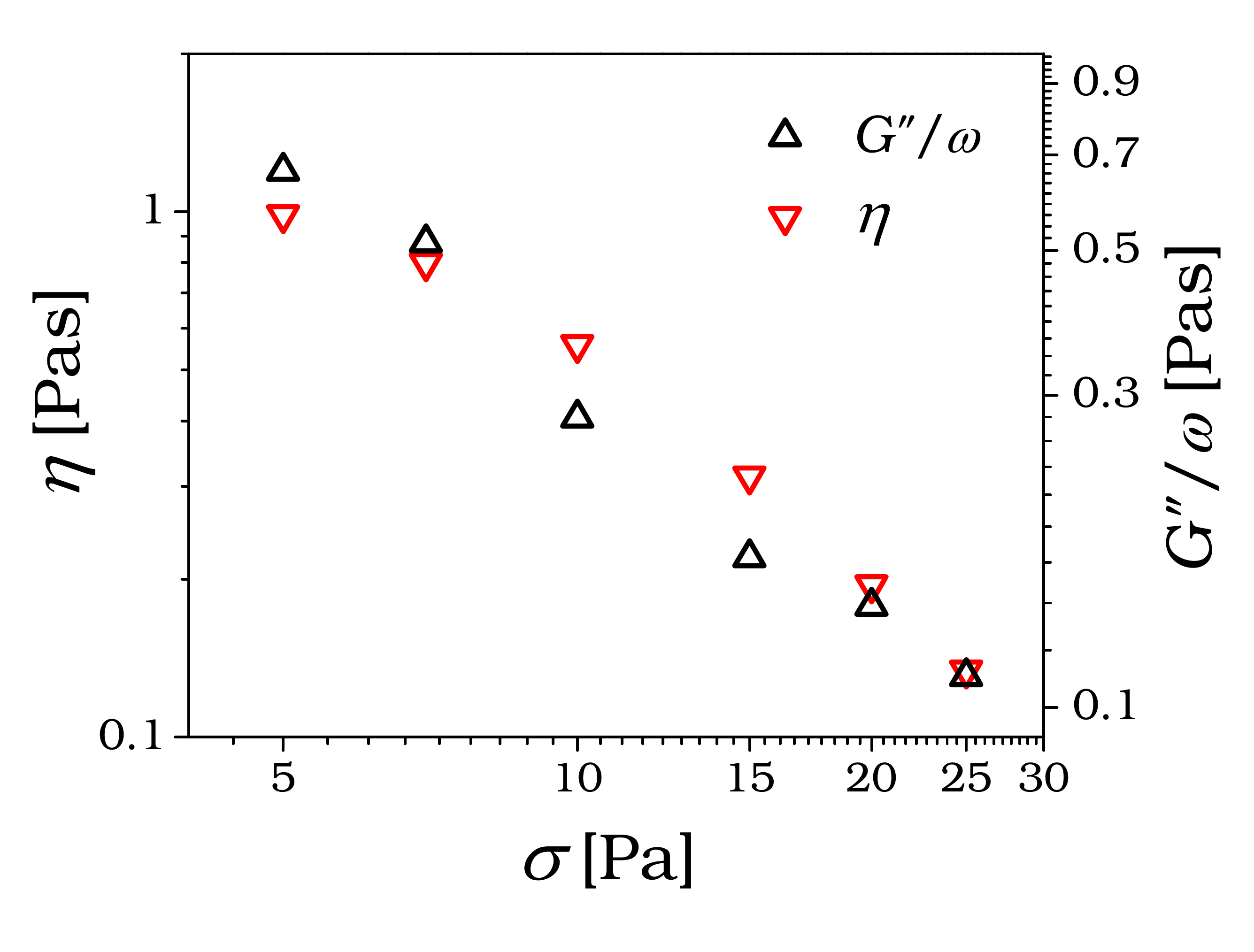}
  \caption{Initial $G''/\omega$ for an oscillatory test (angular
    frequency $\omega $= 0.628 rad/s) plotted as a function of
    magnitude of stress for 2.8 weight \% 9 days old Laponite
    suspension. The viscosity measured from a stress controlled shear
    experiment is also plotted against stress for the same system.}
  \label{fgr:appendixfigure}
\end{figure}

For experiments carried out on 2.8 weight \% Laponite suspension on
day 9, the initial $G''$, measured in the first cycles of oscillatory
shear after shear melting, was found to decrease with stress
amplitude. Since all the samples have equivalent histories prior to
this point, we believe that this decrease represents shear thinning of
the rejuvenated suspension, not a stress-dependent initial solid
fraction. To verify this we measured viscosity ($\eta$) of a sample
immediately after the rejuvenation stage. The
figure~\ref{fgr:appendixfigure} shows $G''/\omega$ and $\eta$ vs
stress. Both the variables show the same dependence on shear stress,
confirming that the decrease in $G''$ is indeed due to shear thinning
of the rejuvenated suspension.

\subsection*{B. Modified exponents for power law traps with variable
  local elastic constant}

Here we consider conditions where $k=k(E)$, so that the elastic
modulus of a trap depends on its depth. For simplicity we assume
$k(E)=E^{p}$.  Equation (2) for the liquid-solid conversion factor $R$
is unaffected, where now $E_{0}/k(E_{0})=\gamma_{0}^2/2$, so that
$\gamma_{0} \propto E_{0}^{(1-p)/2}$. Also the modulus of the solid
material $G$ in equation (3) now obeys:

\begin{equation}
  G \propto \frac{\int\limits_{E_{0}} ^\infty \rho (E) k (E)\,\mathrm{d}E}{\int\limits_{E_{0}} ^\infty \rho (E)\,\mathrm{d}E} \propto E_{0}^{p}
\end{equation}

In other words, the solid fraction populates wells of depth $E> E_{0}$
with the prior distribution, and the modulus of the solid phase is
fixed by the appropriate weighted average of the elastic constants $k$
of the individual wells. This gives $S \propto \sigma
E_{0}^{-(1-p)/2}$, while $R \propto E_{0}^{1-y}$ as before, and
therefore $\theta_{c} \propto \epsilon^{(3-2y+p)/(1+p)}
\sigma^{2(y-1)/(1+p)}$.

\end{document}